\documentclass[useAMS,usenatbib]{mn2e}

\usepackage{graphicx}

\newcommand{\ion}[2]{#1\,{\sc #2}}

\title[Short time scale spectral variability in HD\,92207]{
Short time scale spectral variability in the A0 supergiant HD\,92207 and
the importance of line profile variations
for the interpretation of FORS\,2 spectropolarimetric observations\thanks
{Based on data obtained at the European Southern Observatory (ESO 
Prgs.\ 078.D-0330(A), 87.D-0049(A) and 088.D-0425(A)).}}

\author[Hubrig et al.]{S. Hubrig$^{1}$\thanks{E-mail: shubrig@aip.de},
M. Sch\"oller$^{2}$,
A.~F.~Kholtygin$^{3,4}$\\
$^1$ Leibniz-Institut f\"ur Astrophysik Potsdam (AIP), An der Sternwarte 16, 14482 Potsdam, Germany\\
$^2$ European Southern Observatory, Karl-Schwarzschild-Str.~2, 85748~Garching, Germany\\
$^3$ Astronomical Institute, Saint-Petersburg State University, Saint-Petersburg, Russia \\
$^4$ Isaac Newton Institute of Chile, Saint-Petersburg Branch, Russia
}

\begin{document}

\date{Accepted Received; in original form}

\pagerange{\pageref{firstpage}--\pageref{lastpage}} \pubyear{2010}

\maketitle

\label{firstpage}

\begin{abstract}

Our recent search for the presence of a magnetic field in 
the bright early A-type supergiant HD\,92207 using FORS\,2 in spectropolarimetric mode 
indicated the presence of a longitudinal magnetic field of the order of a few 
hundred Gauss.
Assuming the ideal case of a non-variable star, this discovery has recently been questioned in one work
trying to demonstrate the importance of non-photon noise in FORS\,2 observations.
The assumption of non-variability of HD\,92207 can, however, not be held since substantial profile
variations of diverse lines on a time scale of minutes  or maybe even a fraction of a minute are detected in FORS\,2 spectra.
The presence of short-term spectral variability in blue supergiants, which are considered as type~II 
supernova progenitors, has not been a subject of 
systematic studies before and is critical for the current theoretical 
understanding of their physics. Given the detected short term variability, the question of the presence 
of a magnetic field
cannot be answered without proper modeling of the impact of such a variability on the 
measurements of the magnetic field.
Since the short-term periodicity does not fit into the currently known
domain of non-radially pulsating supergiants, its confirmation is of great importance for 
models of stellar evolution.

\end{abstract}


\section{Introduction}
\label{sect:intro}

Recent developments in observational techniques and theories reveal the potential significance of 
magnetic fields for stellar structure, evolution, and circumstellar environment. At present, 
the distribution of magnetic field strengths in massive stars from the zero-age main sequence 
to more evolved stages, which would shed light on the origin of the magnetic field, has not been 
systematically studied. Our recent search for the presence of a magnetic field in 
the visually brightest early A-type supergiant HD\,92207, using FORS\,2 in spectropolarimetric mode,
resulted in the discovery of a rather strong mean longitudinal magnetic field of the order of a few 
hundred Gauss \citep{Hubrig2012}.

The photometrically and spectroscopically variable bright A0 supergiant star HD\,92207 is of 
particular interest for spectropolarimetric studies. It has been monitored for several years in 
the $uvby$-Str\"omgren system by \citet{Sterken1983} and spectroscopically by 
\citet{Kaufer1996,Kaufer1997}, 
who found cyclical changes of the brightness and substantial profile changes for metal lines and at 
H$\alpha$, and suggested that the observed photometric and H$\alpha$ line variations are the result 
of a corotating structure in the wind, which they considered to be in the star's equatorial plane. 
Furthermore, their study of the line profile variations revealed clear pulsation-like structures, 
indicating the presence of non-radial pulsations (NRPs) with a period of 27\,days, while the stellar 
rotation period is of the order of several months.

\citet{Ignace2009} measured linear polarisation 
in the spectra of this star on seven different nights, 
spanning approximately three months in time. For the continuum polarisation, the 
spiral-shaped wind density enhancement in the equatorial plane of the star suggested by 
\citet{Kaufer1996} was explored.
Importantly, the authors reported that the polarisation across 
the H$\alpha$ line on any given night is typically different from the degree and position angle 
of the polarisation in the continuum. These night-to-night variations in the H$\alpha$ polarisation 
are hard to understand in terms of the spiral structure that was considered for the continuum polarisation. 

Recently, \citet{Bagnulo2013} claimed that the discovery of a longitudinal magnetic field in 
FORS\,2 data is spurious due to non-photon noise, more specifically due to small offsets in the 
parallel and perpendicular beams, or non-predictable instrument instabilities or flexures,
evidenced by changes in the individual spectra.
In all considerations of possible culprits playing a role in the 
magnetic field determination, Bagnulo et al.\ explicitly exclude the role of intrinsic
spectral variations, assuming the ideal
case of a non-variable star, referring to HARPSpol observations obtained in 2013.

In this study, we present a careful inspection of the FORS\,2 spectra used for the 
magnetic field determination in HD\,92207 in our previous work. 
We report on the detection of short-term variability in this object, implying that the 
assumption of an ideal case of a non-variable star cannot be held.

\section{The presence of short-term spectral variations in the A0 supergiant HD\,92207}
\label{sect:descr}

\begin{table}
\caption[]{
Dates of spectropolarimetric observations and radial velocity measurements in the FORS\,2 spectra of HD\,92207.
}
\label{tab:var_meas}
\centering
\begin{tabular}{lr@{$\pm$}lr}
\hline
\hline
\multicolumn{1}{c}{MJD} &
\multicolumn{2}{c}{RV} &
\multicolumn{1}{c}{S/N} \\
\multicolumn{1}{c}{} &
\multicolumn{2}{c}{[km/s]} &
\multicolumn{1}{c}{} \\
\hline
55688.168 & $-$38.19 & 9.50 & 2020 \\
55936.341 & $-$68.86 & 9.62 & 1942 \\
56018.224 & $-$12.36 & 8.07 & 1783 \\
\hline
\end{tabular}
\end{table}

The FORS\,2 spectropolarimetric observations on three different epochs discussed in this work have been 
used by \citet{Hubrig2012}
for a search of the presence of a magnetic field in the bright A0 supergiant HD\,92207. The dates of observations,
the radial velocities measured in the obtained spectra, and their S/N are presented in Table~\ref{tab:var_meas}.

Apart from magnetic field measurements, \citet{Hubrig2012} studied 
the spectrum variability over the last seven
years, adding the seven FORS\,1 observations obtained in 2007, one FEROS spectrum 
($R=\lambda/\Delta\lambda\approx48\,000$) observed 
in 2004 on La Silla (retrieved from the ESO archive), and one EBASIM spectrum 
($R=\lambda/\Delta\lambda\approx20\,000$) obtained in 2006 in CASLEO in Argentina.
Distinct line profile 
variations of all spectral lines, which are caused by non-radial pulsations (NRPs) with a period of 27\,days
\citep{Kaufer1997},
have been detected in all observations. Furthermore, \citet{Hubrig2012} have reported that the 
line profiles of the elements Fe, Cr, and Si show very similar behaviour.
The radial velocities determined from the 
\ion{He}{i} $\lambda$ 5876 line are only slightly higher than those for the metal lines, 
while the Balmer line H$\beta$ shows
a velocity that is systematically lower than the values obtained from the metal lines.

Assuming the presence of an effect of the beam-swapping technique on wavelength shifts and the possible occurrence of 
offsets due to a sporadic instability of the instrument, \citet{Bagnulo2013} argue that no 
magnetic field is present in the atmosphere of HD\,92207, calling the reader's attention to the 
important role of non-photon noise in the spectropolarimetric observations with FORS\,2.
In all considerations of possible culprits playing a role in the 
magnetic field determination, Bagnulo et al.\ assume the ideal
case of a non-variable star referring to HARPSpol observations of HD\,92207 obtained in 2013.
These observations have exposure times
between 30 and 50\,min and have been used to construct LSD profiles, which combine hundreds of spectral lines of 
various elements under the assumption that all spectral lines are identical in shape and can 
be described by a scaled mean profile. 
So far, the published analyses \citep{Bagnulo2013,Hubrig2012} of spectropolarimetric observations
of HD\,92207, either with HARPS or FORS, have not taken into account the effect of short-term
line variability.
Also the current analysis is not able to properly address this variability.

A detailed description of the assessment of the longitudinal 
magnetic field measurements using FORS\,1/2 is presented in our previous work
(e.g.\ \citealt{Hubrig2004a, Hubrig2004b}, and references therein). 
The mean longitudinal 
magnetic field, $\left< B_{\rm z}\right>$, is derived using 

\begin{equation} 
\frac{V}{I} = -\frac{g_{\rm eff} e \lambda^2}{4\pi{}m_ec^2}\ \frac{1}{I}\ 
\frac{{\rm d}I}{{\rm d}\lambda} \left<B_{\rm z}\right>, 
\label{eqn:vi}
\end{equation} 

\noindent 
where $V$ is the Stokes parameter that measures the circular polarisation, $I$ 
is the intensity in the unpolarised spectrum, $g_{\rm eff}$ is the effective 
Land\'e factor, $e$ is the electron charge, $\lambda$ is the wavelength, $m_e$ the 
electron mass, $c$ the speed of light, ${{\rm d}I/{\rm d}\lambda}$ is the 
derivative of Stokes~$I$, and $\left<B_{\rm z}\right>$ is the mean longitudinal magnetic 
field. 

\begin{figure}
\centering
\includegraphics[angle=270,width=0.45\textwidth]{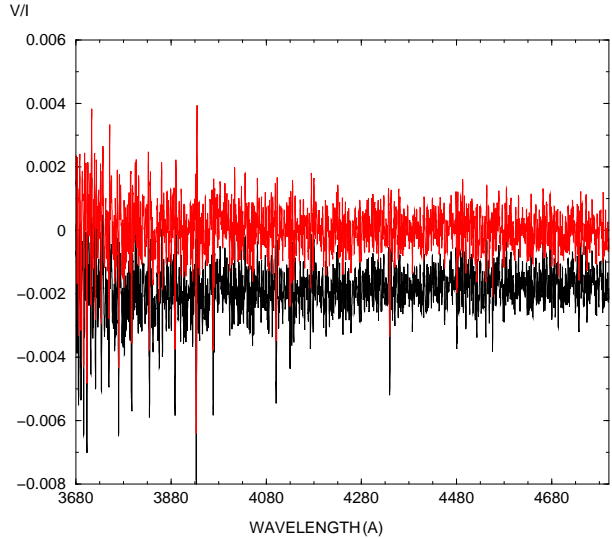}
\caption{
Rectified (red) and non-rectified (black) Stokes $V/I$ spectra of HD\,92207 observed on the first epoch in May 2011.
}
\label{fig:rectif}
\end{figure}

This field is usually measured in two ways: using only the absorption hydrogen Balmer
lines or using the entire spectrum including all available absorption lines in spectra
that were rectified. 
The impact of rectification is usually very weak:
for HD\,92207 the difference between the mean longitudinal magnetic field obtained 
in rectified spectra and that obtained 
in non-rectified spectra is only about 15\,G.
Even for the highly linearly polarized object Cyg\,X--1, \citet{Bochkarev2012} report only
a difference between the rectified and the non-rectified Stokes~$V$ spectra
of up to 20\,G in the FORS\,1 observations.
The rectified and non-rectified $V/I$ spectra of HD\,92207 observed on the first epoch in May 2011 are presented
in Fig.~\ref{fig:rectif}.

Two spectral regions were excluded from the measurements of Hubrig et al.\ (2012): one is related 
to the spectral region containing the H$\beta$ line because the high-resolution spectra of this star
show a contamination of this line by emission. The second spectral region excluded from the 
measurements is the region 3957--3975\,\AA{}, where
H$\epsilon$ is blended with the \ion{Ca}{ii}\,H line.
In addition, an inspection of the spectra obtained with the blue optimised chip
reveals the presence of instrumental artifacts at wavelengths close to the \ion{Ca}{ii} K and 
H$\epsilon$ lines.

To minimise the cross-talk effect, a sequence of subexposures at the retarder position 
angles $-45^{\circ}+45^{\circ}$, $+45^{\circ}-45^{\circ}$, $-45^{\circ}+45^{\circ}$, etc.\ is usually 
executed during observations and the values $V/I$ are calculated using:

\begin{equation}
\frac{V}{I} = \frac{1}{2} \left\{ \left( \frac{f^{\rm o} - f^{\rm e}}{f^{\rm o} + f^{\rm e}} \right)_{-45^{\circ}}
                               - \left( \frac{f^{\rm o} - f^{\rm e}}{f^{\rm o} + f^{\rm e}} \right)_{+45^{\circ}} \right\}
\end{equation}

\noindent
where $+45^{\circ}$ and $-45^{\circ}$ indicate the position angle of the retarder waveplate and $f^{\rm o}$ and $f^{\rm e}$ are the ordinary and 
extraordinary beams, respectively.
In the calculations, a Land\'e factor $g_{\rm eff}=1$ for the hydrogen lines 
and $g_{\rm eff}=1.25$ for the metal lines is assumed. 
For the first observational epoch, \citet{Hubrig2012} collected eight subexposures, each one with
an exposure time of three seconds, recorded at time intervals between 1.6--2.1\,minutes,
while for each the second and the third epochs the collected number of subexposures was twelve.

\begin{figure}
\centering
\includegraphics[width=0.45\textwidth]{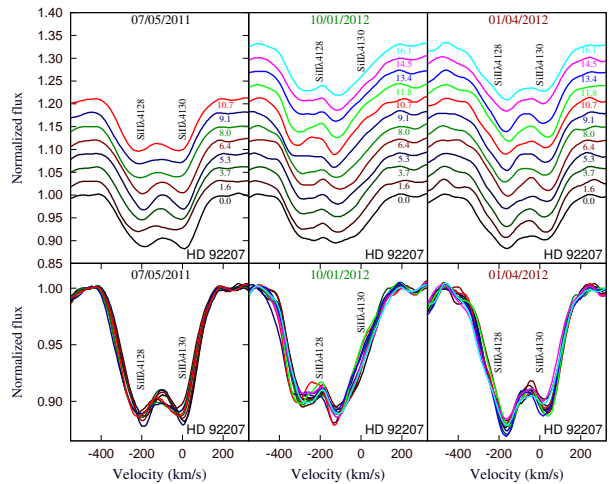}
\caption{The behaviour of the \ion{Si}{ii} doublet $\lambda\lambda$4129, 4131
in the FORS\,2 spectra 
of HD\,92207 in each individual subexposure belonging to the observations on the three different epochs.
For each epoch, in the upper row, we present the line profiles shifted in vertical
direction for best visibility. The time difference (in minutes) between the subexposure and the
start of observations is given close to each profile. The lower row shows all profiles overplotted.
Distinct profile variations are detected already on time scales of minutes.
}
\label{fig:si}
\end{figure}

\begin{figure}
\centering
\includegraphics[width=0.45\textwidth]{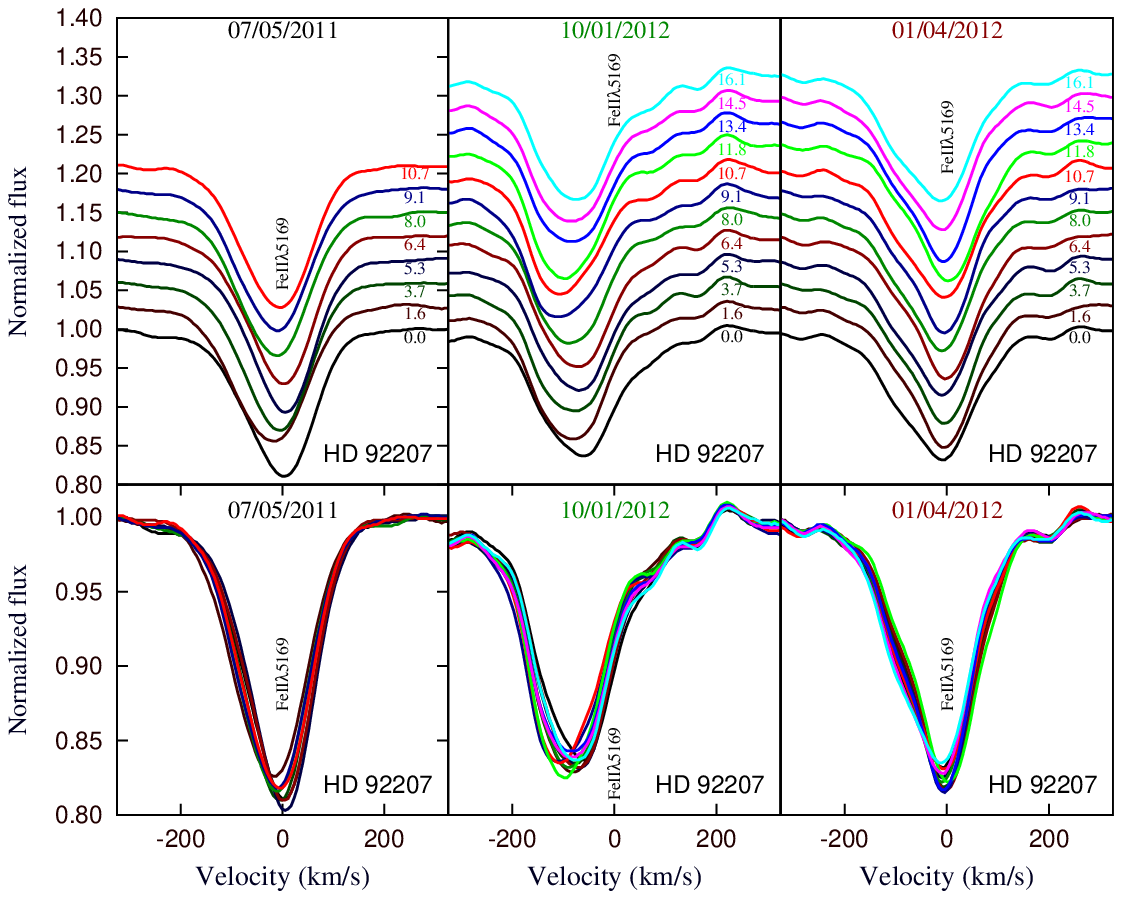}
\caption{ The behaviour of the \ion{Fe}{ii} $\lambda$5169 line in the FORS\,2 spectra 
of HD\,92207 in each individual subexposure belonging to observations on three different epochs.
}
\label{fig:fe}
\end{figure}

\begin{figure}
\centering
\includegraphics[width=0.45\textwidth]{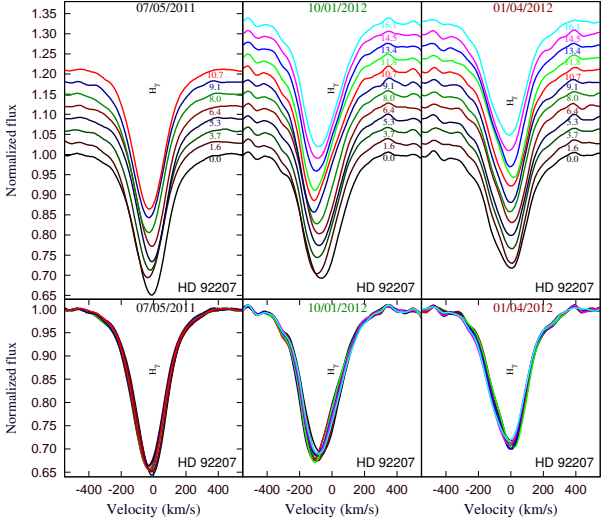}
\caption{The behaviour of the H$\gamma$ line in the FORS\,2 spectra of HD\,92207 in each individual subexposure 
belonging to observations on three different epochs.
}
\label{fig:gamma}
\end{figure}

\begin{figure}
\centering
\includegraphics[width=0.40\textwidth]{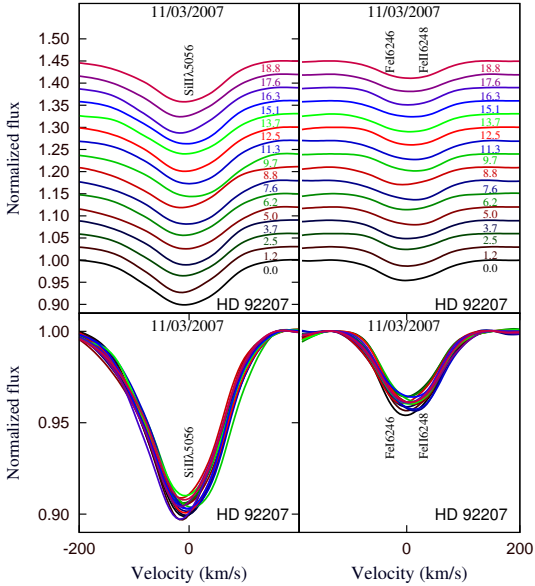}
\caption{
The behaviour of the \ion{Si}{ii}\,$\lambda$5056 and \ion{Fe}{ii}/\ion{Fe}{i}\,$\lambda$6246--6247 lines
in the FORS\,1 spectra 
of HD\,92207 in each subexposure, recorded at time intervals between 1.2 and 1.4\,minutes.
}
\label{fig:fors1}
\end{figure}

The inspection of the FORS\,2 spectra obtained in each individual subexposure 
reveals distinct line profile
variations not only between the different observation epochs, but also on a time scale of a couple of minutes. 
Along with different radial velocity 
shifts detected for lines belonging to different elements, we also detect notable changes in
line intensities taking place on very short time scales. In Figs.~\ref{fig:si}, \ref{fig:fe}, and
\ref{fig:gamma}, we present 
the clearly visible variations
of line profiles belonging to different elements in individual subexposures reaching up to 3\% in 
intensities and up to 30\,km\,s$^{-1}$ in radial velocities.
Notably, the  H$\gamma$ line profile presented in Fig.~\ref{fig:gamma}
exhibits intensity and radial velocity variations in individual subexposures at a lower level 
compared to the metal lines. 
The short-term spectral variability of HD\,92207 is also detected in earlier spectropolarimetric observations 
carried out with FORS\,1 in 2007.
These observation were obtained with exposure times of about 5\,s to study linear polarization across
the H$\alpha$ profile, to associate the detected changes with a corotating structure \citep{Ignace2009}.
As in observations obtained  with FORS\,2, conspicuous line profile variations with similar 
amplitudes in radial velocities and line intensities are clearly visible  
in the FORS\,1 spectra.  As an example, we present in Fig.~\ref{fig:fors1} the variability of the
Si and Fe line profiles.
 
\begin{figure}
\centering
\includegraphics[width=0.45\textwidth]{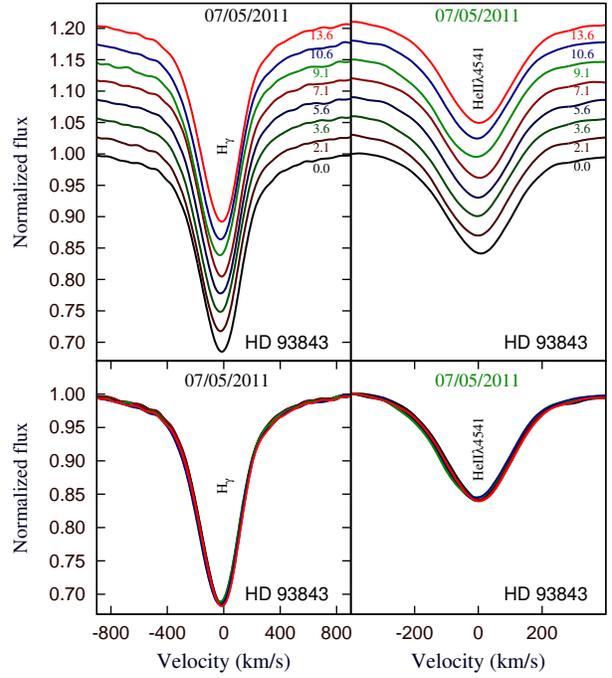}
\caption{
The behaviour of the H$\gamma$ and \ion{He}{ii}\,$\lambda$4541 lines in the FORS\,2 spectra 
of HD\,93843 in each subexposure,
obtained immediately after the HD\,92207 observations in the first epoch.
}
\label{fig:hd93}
\end{figure}

Immediately after completing the observations for 
HD\,92207 with FORS\,2 in visitor mode in 2011 May, \citet{Hubrig2012} observed 
close to the position of HD\,92207 the hot massive star 
HD\,93843 at the same air mass and with exposure times of 30\,s.
As is shown in Fig.~\ref{fig:hd93}, opposite to the spectral behaviour 
of HD\,92207, no significant variability in intensity, i.e.\ only intensity variations below 1\% 
and radial velocities below 10\,km\,s$^{-1}$ have been detected in the individual subexposures 
obtained for this star.

\begin{figure}
\centering
\includegraphics[width=0.45\textwidth]{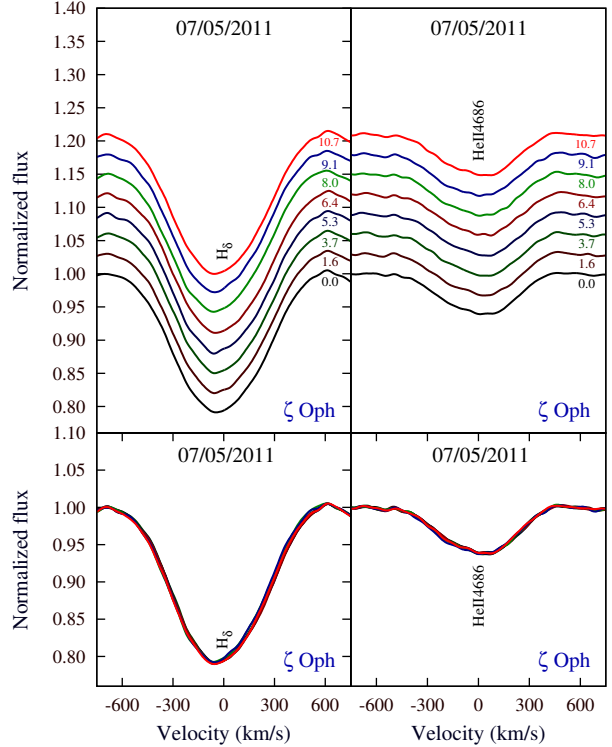}
\caption{
The behaviour of the H$\delta$ and \ion{He}{ii}\,$\lambda$4686 lines in the FORS\,2 spectra 
of $\zeta$\,Oph in each subexposure,
obtained 1.5\,h after the HD\,92207 observations in the first epoch.
}
\label{fig:zeta}
\end{figure}

Furthermore, we inspected in addition the behaviour of line profiles in the FORS spectra of 
another bright star, $\zeta$\,Oph, observed with an exposure time of 0.2\,s in the same night in the same weather 
conditions just 1.5\,h after the observations of HD\,92207.  Also in these observations 
no significant variability, i.e.\ only intensity 
variations below 1\% and radial velocities below 10\,km\,s$^{-1}$,
have been detected in the individual subexposures. 
We display the behaviour of the H$\delta$ and \ion{He}{ii}\,$\lambda$4686 lines in the FORS\,2 spectra 
of $\zeta$\,Oph in each subexposure in Fig.~\ref{fig:zeta}.

\begin{figure}
\centering
\includegraphics[width=0.35\textwidth]{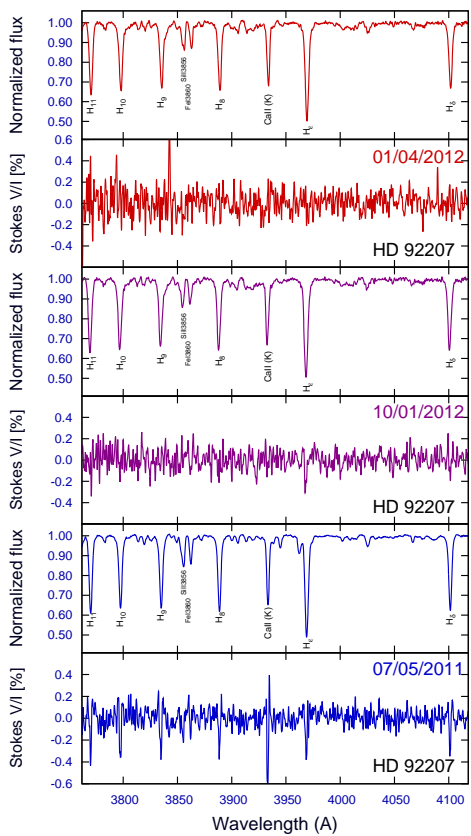}
\caption{
Stokes $I$ and $V$ spectra of HD\,92207 in a spectral region containing low-number Balmer lines
 and the Ca doublet lines at three epochs. 
}
\label{fig:magn}
\end{figure}

\begin{figure}
\centering
\includegraphics[width=0.335\textwidth]{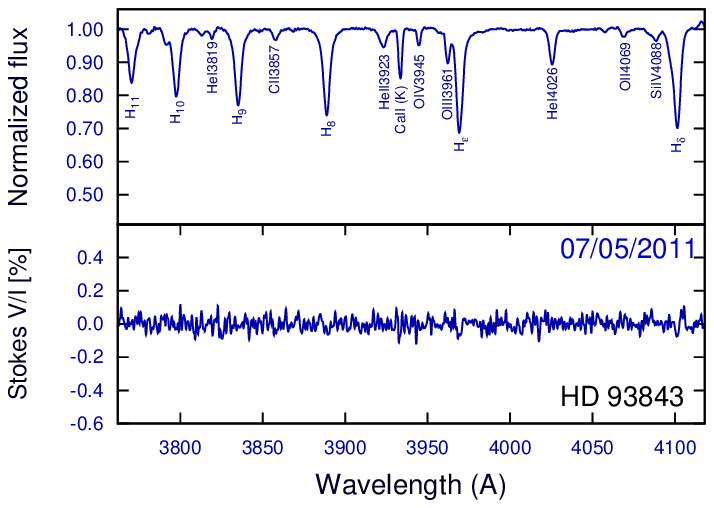}
\includegraphics[width=0.335\textwidth]{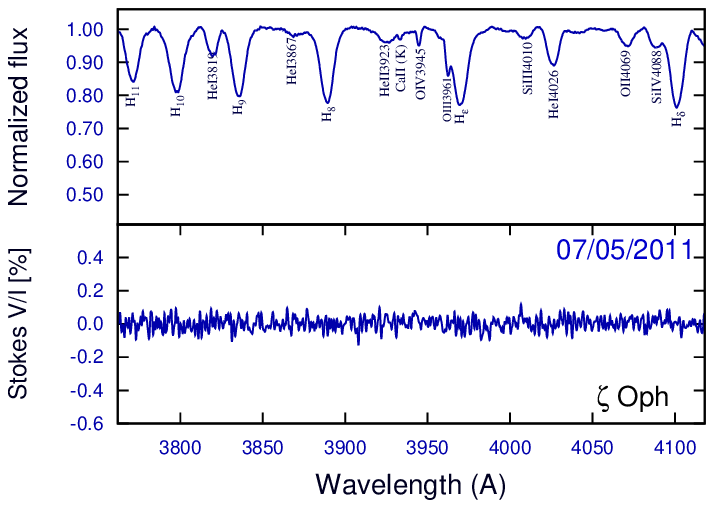}
\caption{
Stokes $I$ and $V$  spectra in the same spectral region as in Fig.~\ref{fig:magn} during the first epoch in May 2011.
{\sl Top:} The massive O-type star HD\,93843.
{\sl Bottom:} The massive O-type star $\zeta$\,Oph.
}
\label{fig:magn2}
\end{figure}

The absence of strong spectral variability in HD\,93843 and $\zeta$\,Oph on similar time scales 
suggests that the spectral variations of HD\,92207 are intrinsic and are not related to an 
imperfect performance of FORS\,2.
If non-photon noise was indeed present in the FORS\,2 spectropolarimetric observations of that night and the 
short term variability detected in the spectral lines of HD\,92207 had no effect on
the measurement of the magnetic field, we would correspondingly 
expect the detection of a weak magnetic field in HD\,93843 and $\zeta$\,Oph.

In Fig.~\ref{fig:magn}, we present Stokes $I$ and $V$ spectra of HD\,92207 in a spectral region containing 
low-number Balmer lines and the Ca doublet lines at three epochs.
As can be seen, the Zeeman features corresponding to several spectral lines in the lowest panel  do not 
exhibit the classical ``S'' shape, but instead show a deep negative polarization trough without the 
equivalent positive polarization spike.
While \citet{Bagnulo2013} explain those by various offsets due to instrumental misbehaviour,
they are in fact very likely resulting from the spectral variations in this object.
However, as we show in Fig.~\ref{fig:magn2}, 
no similar Zeeman features
are observed in the spectra of HD\,93843 and $\zeta$\,Oph, with measured mean longitudinal magnetic fields 
$\left<B_{\rm z}\right> = -63\pm47$\,G, and $\left<B_{\rm z}\right> = 95\pm68$\,G, respectively \citep{Hubrig2013}.
These measurements are compatible with zero.
Therefore, we conclude that the effects discussed by 
\citet{Bagnulo2013} are of no importance for the spectropolarimetric observations of HD\,92207 and that
the discovered line shifts in the polarimetric 
spectra of HD\,92207 are simply explained by the variability of this star on a short time scale and which
can not be found in HD\,98343 and $\zeta$\,Oph.

\section{Reassessing our earlier magnetic field determination}
\label{sect:mf}

\begin{table}
\caption[]{
Reassessment of the magnetic field measurements of HD\,92207.
}
\label{tab:mfs_meas}
\centering
\begin{tabular}{lr@{$\pm$}lrr@{$\pm$}l}
\hline
\hline
\multicolumn{1}{c}{MJD} &
\multicolumn{2}{c}{$\left<B_{\rm z}\right>_{\rm LR}$} &
\multicolumn{1}{c}{$\sqrt{\chi^2_{\rm min}/\nu}$} &
\multicolumn{2}{c}{$\left<B_{\rm z}\right>_{\rm BS}$} \\
\multicolumn{1}{c}{} &
\multicolumn{2}{c}{[G]} &
\multicolumn{1}{c}{} &
\multicolumn{2}{c}{[G]} \\
\hline
 55688.168 & $-$355 & 76 & 0.80 & $-$355 & 99 \\
 55936.341 &    147 & 62 & 1.17 &    147 & 62 \\
 56018.224 &     60 & 76 & 1.07 &     60 & 81 \\
\hline
\end{tabular}
\end{table}

We have reassessed our magnetic field determinations, which we present
in Table~\ref{tab:mfs_meas}.
The first column gives the modified Julian date at the middle of the exposure
and in the second column, we list the magnetic field and its error determined
by the usual linear regression method.
In the following two columns we first show the $\sqrt{\chi^2_{\rm min}/\nu}$
value, which is an indicator for the quality of the assumed errors and the
validity of Eq.~\ref{eqn:vi},
and the magnetic field and its error determined from our Monte Carlo bootstrapping tests.
These latter two indicators are described in detail below.
The longitudinal magnetic field was for all three epochs determined from data
in the wavelength regions 3250--3957\,\AA{} and 3975--4855\,\AA{},
i.e.\ we are not using H$\beta$, H$\epsilon$, or \ion{Ca}{ii}\,H.
In the measurements of \citet{Hubrig2012}, we also used
the wavelength region beyond H$\beta$, which explains the now larger errors.

Before the linear regression, we performed a rectification of the $V/I$ spectra.
This step was also performed in \citet{Hubrig2012}, which should
be obvious from Fig.~1 in that paper.
For the rectification, we fit and subtract a line to the data.
This is somewhat similar to one of the two methods proposed by \citet{Bagnulo2012},
who fit a ``smooth function'', which we assume is a low order polynomial.
However, we do not see the need for a higher order than a linear fit.
We also would like to note that only the slope of this line is relevant for the determination
of the longitudinal magnetic field.
As already discussed in Sect.~\ref{sect:descr}, the influence of the rectification on
the magnetic field determination is well within the error bars.

We are interpolating all FORS\,2 spectra, leading to a spectral bin size of 0.1\,\AA{},
compared to the natural FORS\,2 spectral bin size of 0.75\,\AA{}.
This way, we are generating new data, which can not be treated as individual
measurements.
Thus, we have to take a factor of $\sqrt{7.5}$ into account when calculating the errors
both for the standard linear regression and in the Monte Carlo bootstrapping analysis.
This factor is not relevant for the determination of $\sqrt{\chi^2_{\rm min}/\nu}$.

In the analysis of high-resolution spectropolarimetric observations,
null polarisation spectra are frequently calculated to assure that no 
instrumental or data-reduction effects are present in the measurements
of magnetic fields. They are calculated by combining the sub-exposures
in such a way that the polarisation cancels out.
Our experience with the 
calculation of null profiles for FORS\,2 polarimetric observations shows that 
they cannot be considered as reliable indicators for the presence or non-presence of magnetic fields 
since the null profiles obtained by sorting the frames in different ways are different. 
Also, signals in the null profile might be the result of combining data with
significantly different signal-to-noise (Sch\"oller et al., in prep.).
\citet{Bagnulo2012} suggest to calculate an ``external'' error bar for the
magnetic field values determined in the linear regression by multiplying the internal error
with the factor $\sqrt{\chi^2_{\rm min}/\nu}$,
where $\nu$ is the number of degrees of freedom of the system, i.e.\ the number
of spectral bins minus 2, and $\chi^2_{\rm min}$ is calculated from 

\begin{equation}
\chi^2_{\rm min} = \sum_i \frac{(y_i-\left<B_{\rm z}\right>x_i-b)^2}{\sigma^2_i}
\end{equation}

\noindent
where the sum spans over all spectral bins $i$, $\left<B_{\rm z}\right>$ and $b$
respectively are the longitudinal magnetic field and the instrumental polarisation determined
from the previous linear regression, $y_i$ is the left hand term of Eq.~\ref{eqn:vi} in spectral bin $i$,
and $x_i$ is the factor in front of $\left<B_{\rm z}\right>$ in the right hand term of Eq.~\ref{eqn:vi} in spectral bin $i$.

For the three data sets, we calculate the factor $\sqrt{\chi^2_{\rm min}/\nu}$ presented in Table~\ref{tab:mfs_meas}.
For the first epoch the error seems to be very reasonable, while for the other two epochs we
probably underestimate the errors by 17\% and 7\%, respectively.

\begin{figure*}
\centering
\includegraphics[width=0.32\textwidth]{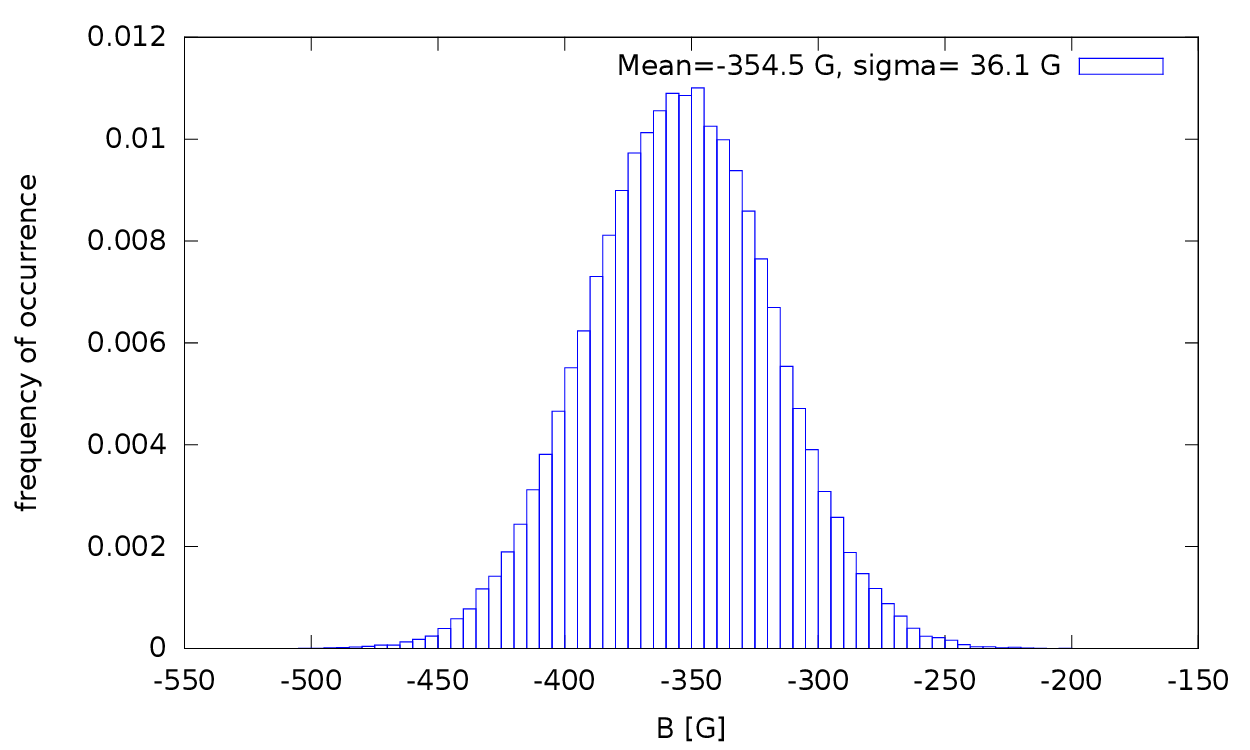}
\includegraphics[width=0.32\textwidth]{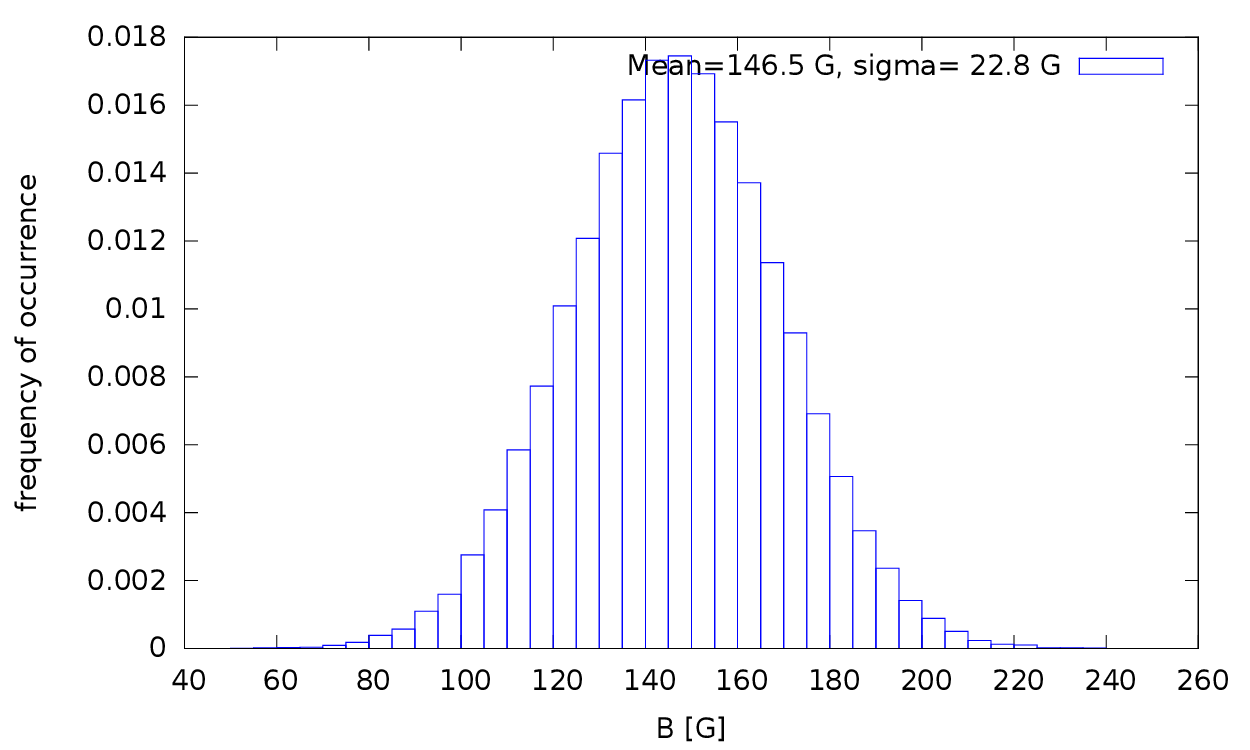}
\includegraphics[width=0.32\textwidth]{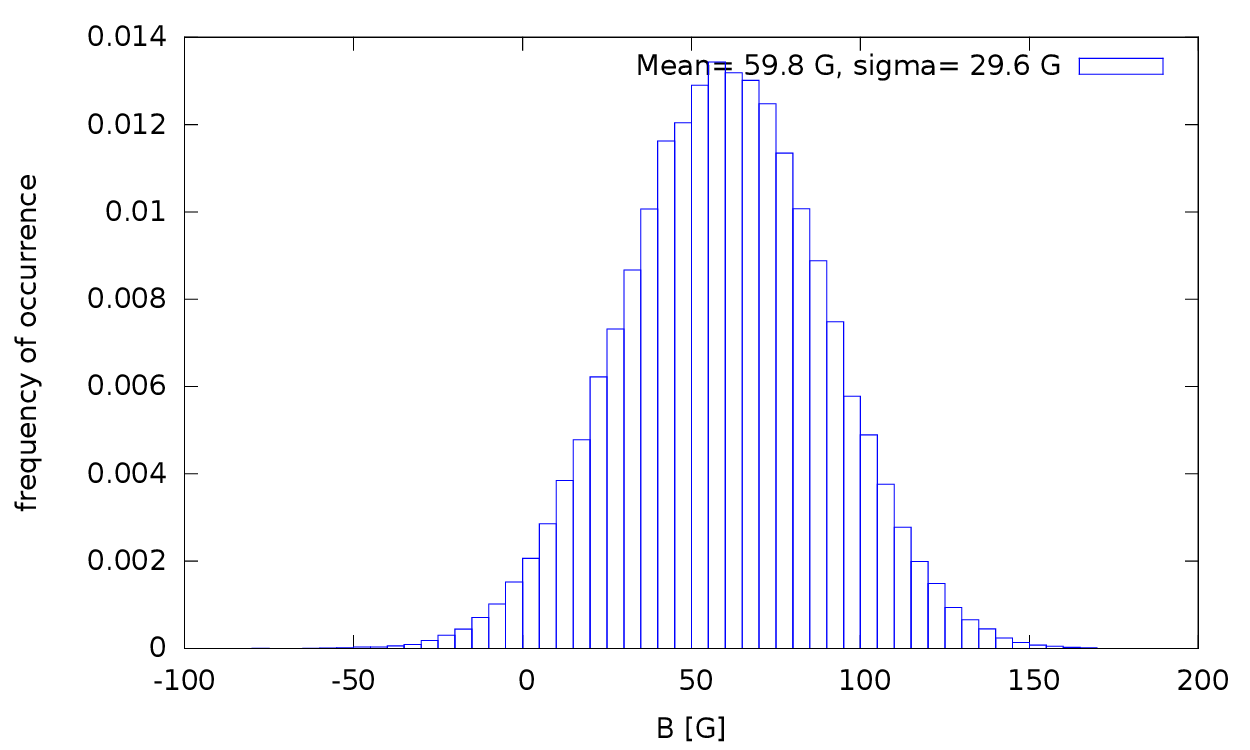}
\caption{
Distributions from our Monte Carlo bootstrapping tests for the data sets of
HD\,92207 from the three different epochs.
The widths of these profiles are used to determine the errors listed in Table~\ref{tab:mfs_meas}.
}
\label{fig:boot}
\end{figure*}

Following the suggestion by \citet{Rivinius2010}, we also carried out Monte Carlo bootstrapping tests.
These are most often applied with the purpose of deriving robust 
estimates of standard errors.
In these tests, we generate 250\,000 samples that have the same size as the original data set (15\,866 spectral bins)
and analyse the distribution of the $\left<B_{\rm z}\right>$ determined from all these newly generated data sets.
The resulting distributions, which we present in Fig.~\ref{fig:boot}, all show smooth Gaussian profiles.

From this reassessment, we conclude that our data reduction does not show any flaw.
However, given the detected short term variability, the question of the presence of a magnetic field
cannot be answered without proper modeling of the impact of such a variability on the 
measurements of the magnetic field. High resolution spectropolarimetric observations obtained on a short time scale are 
urgently needed before any conclusion can be drawn.

\section{Discussion}
\label{sect:disc}

The main conclusion by \citet{Bagnulo2013} on the discrepancies between their and our earlier
determination of the longitudinal magnetic field seems to be a wrong relative wavelength 
calibration on our side, especially due to the impact of wavelength shifts when
rotating the retarder waveplate.
While the assessment of a potential influence of a wrong relative wavelength
calibration seems to be correct, it is not plausible that this is actually
an issue for the data from the first epoch, where the longitudinal magnetic field is detected.
Directly after the observations of HD\,92207, we observed the hot massive star HD\,93843
close to the position of HD\,92207, at the same air mass and with similar short exposure times.
We do not find any distinct Zeeman features for HD\,93843 and the determined longitudinal
magnetic field is compatible with zero.
HD\,93843 also does not show strong spectral variability on a short time scale, ruling
out that spectral variability is introduced due to imperfections within FORS\,2.
The same results are obtained from observations of $\zeta$\,Oph, with an exposure time
of 0.2\,s.
In the same night, we also observed the two Of?p stars HD\,148937 and CPD\,$-$28\,2561 \citep{Hubrig2013}.
The measurement of a magnetic field of $139\pm33$\,G in HD\,148937 using the same wavelength calibration
is perfectly in line with earlier
measurements by ourselves \citep{Hubrig2011a,Hubrig2013} and \citet{Wade2012}, following the 7.032\,d period
determined by \citet{Naze2010}.
For CPD\,$-$28\,2561, we find a magnetic field of $269\pm81$\,G, which is supported by
\citet{Petit2013}, who give a polar field strength larger than 1.7\,kG for CPD\,$-$28\,2561.
All these measurements indicate that there is nothing wrong with the wavelength calibration
for that night.
This still leaves room for a spontaneous instability in the FORS\,2 wavelength calibration
that affects the observations of HD\,92207 but no other object.
While we can not rule this out, there is no evidence that could support this speculation
in this or any other FORS\,1/2 data set.
Another point raised by \citet{Bagnulo2013} was that we could underestimate our magnetic field errors.
We have used the mechanism to estimate the ``external'' error, as described by
\citet{Bagnulo2012} and can conclude that there is also no issue with
the data in that respect.
We also used the Monte Carlo bootstrapping tests proposed by \citet{Rivinius2010},
which indicate that we only slightly understimated our initial errors.
We suggest that what \citet{Bagnulo2013} claim to result from non-photon noise,
most probably can be traced back to spectral variability of HD\,92207 on short time scales.

The available FORS\,2 polarimetric spectra clearly show the presence of short-term spectral 
variability, which was not previously discussed in the literature for any blue supergiant and
certainly needs further investigation. In particular, a careful search for periodicity
and identification of pulsation modes causing the remarkable changes in the line intensity and  position
on time scales of the order of minutes are urgently needed. 
With the current data, it can not be decided, if the variations are of
periodic or stochastic nature.
In any case, given the size of the supergiant, it is clear that the variability
can not be referred to coherent line variations across the entire surface on such short time scales.
Obviously, seismic studies are of great importance to constrain physical processes in stars,
e.g. differential rotation, mixing, mass loss, etc..
Despite several decades of observational
efforts with ground-based photometry and spectroscopy of bright blue supergiants,
it is not yet certain what portion of
their variability is periodic, or how far they deviate from strict periodicity. 
On the other hand, in the most recent observational studies of massive stars, evidence is accumulating 
that some BA supergiants exhibit multiperiodic NRPs.
As an example, short-term variability was already identified
on a time scale of 1--3\,hours \citep{Lefever2007,Kraus2012}. 
However, a variability on time scales of the order of minutes has not been detected so far,
mostly due to the fact that telescopes with large collecting areas are needed for studies with 
spectral time sampling of a few minutes.

Clearly, it is not possible to use the low-resolution FORS\,2 spectra to model the effect of
pulsations on the magnetic field measurements, and the potential of high-resolution spectropolarimetric
observations should be used in the search of short-term variations (e.g.\ \citealt{Hubrig2011b}).
We need to note that due to the proprietary time of one year for ESO observations, we are not yet able 
to study the spectral variability of this star in the high resolution HARPSpol spectra mentioned in the work 
of \citet{Bagnulo2013}.
The question how pulsations affect the magnetic field measurements is not yet solved in spite of 
the fact that the number of studies of pulsating $\beta$~Cephei and slowly pulsating B (SPB) stars is 
gradually increasing.
Already in \citeyear{Schnerr2006}, \citeauthor{Schnerr2006} discussed the
influence of pulsations on the analysis of the magnetic field strength in the $\beta$~Cephei star $\nu$~Eri
in MUSICOS spectra and tried to model the signatures found in Stokes $V$ and $N$ spectra. 
Although the authors claim that using some modeling they are able to quantitatively establish
the influence of pulsations on the magnetic field determination, they still detect profiles in
Stokes $N$ and $V$ that are the result of the combined effects of the pulsations and the inaccuracies
in wavelength calibration that were not removed by their imperfect modeling of these effects.

 Another important effect in the measurements of magnetic fields in
 pulsating stars, applying the LSD
 method to high-resolution polarimetric spectra, is that the line profiles belonging to different elements show
 different profile shapes and different displacements.
The authors usually use essentially all metallic lines and He lines
 (up to several hundred lines) to calculate a "mean" LSD line profile, although the behaviour of lines of
 different elements during the pulsation cycle is frequently different (e.g.\ \citealt{Hubrig2011b}).

Blue supergiants are considered as type~II supernova progenitors. A careful study of their variability
provides important diagnostic means for internal and atmospheric structure. The need for multiple 
modes 
to fit to the spectroscopic data sets has already been presented in several works analysing
BA supergiants. According to the study of Rigel by \citet{Moravveji2012},
periods shorter than about a week can only be caused by the $\kappa$-mechanism if other sources 
such as spots, variable winds, and propagating shocks can be excluded.
The goal of future studies should be to search for the presence of short-term variability and  
periodicities in bright A0 supergiants with similar 
stellar parameters.
Moreover, since the short-term periodicity does not fit into the currently known
domain of non-radially pulsating supergiants, its confirmation is of great importance for 
the models of stellar evolution.

\section*{Acknowledgments}

We are grateful to Helge Todt, who provided his Monte Carlo bootstrapping routines.

\label{lastpage}

\end{document}